\documentstyle[epsfig,epsf]{article}
\def\Journal#1#2#3#4{{#1} {\bf #2}, #3 (#4)}

\def\NCB{{\em Nuovo Cimento} B}
\def\NCC{{\em Nuovo Cimento} C}

\def\NIMA{{\em Nucl. Instrum. Methods} A}

\def\PLB{{\em Phys. Lett.}  B}
\def\PRL{\em Phys. Rev. Lett.}
\def\PRD{{\em Phys. Rev.} D}
\def\ZPC{{\em Z. Phys.} C}
\def\NPBPS{{\em Nucl. Phys.} B {\em Proc. Suppl.}}
\def\SJNP{\em Sov. J. Nucl. Phys.}

\begin{document}

\begin{center}
{\Large {\bf ATMOSPHERIC NEUTRINO OSCILLATIONS}}
\end{center}

\vskip .7 cm

\begin{center}
G. GIACOMELLI, M. GIORGINI and M. SPURIO \par~\par
{\it Dept of Physics, Univ. of Bologna and INFN, \\
V.le C. Berti Pichat 6/2, Bologna, I-40127, Italy\\} 

E-mail: giacomelli@bo.infn.it , giorginim@bo.infn.it , spurio@bo.infn.it

\par~\par

Lectures at the 6$^{th}$ School on Non-Accelerator Astroparticle Physics,  
\\ Trieste, Italy, 9-20 July 2001. 

\vskip .7 cm
{\large \bf Abstract}\par
\end{center}

{\normalsize The recent results from the Soudan 2, MACRO and 
SuperKamiokande experiments  
on atmospheric neutrino oscillations are
summarized and discussed. Some features of possible future atmospheric
neutrino experiments are presented.}

\vspace{5mm}

\section{Introduction}
A high energy primary cosmic ray, proton or nucleus, interacts 
in the upper atmosphere producing  a large number of pions 
and kaons, which decay yielding  muons and muon
neutrinos; also the muons decay yielding muon and electron neutrinos.
 The ratio of the
numbers of muon to electron neutrinos is about 2 
and $N_{\nu}/N_{\overline\nu} \simeq 1$.
 These neutrinos are produced in a spherical
surface at few tens of km above ground and  they proceed at 
high speed towards the earth.

The atmospheric neutrino flux was computed by many authors.
 At low energies, $E_\nu \sim 1$ GeV, the numbers of 
neutrinos predicted by different authors differ by about 
$20\div 30\%$ \cite{nulow,Agrawal96};
 at higher energies, $E_\nu > 10$ GeV, the predictions are more
reliable, with an estimated systematic  uncertainty of about 
$15\%$ \cite{Agrawal96,nuhigh}, 
 almost one half of that at low energies. However the predicted
relative rates of $\nu_\mu$ and $\nu_e$ and the shapes of the zenith 
distributions are affected by 
considerably lower systematic errors. Other sources of systematic
uncertainties arise from the knowledge of the 
neutrino-nucleon cross sections and from the propagation of muons and
electrons in different materials.

Several large underground detectors studied atmospheric neutrinos. 
These detectors were and are located below a cover of 1-2 km of rock and 
 may detect 
neutrinos coming from all directions or from below. Via charged current (CC) 
interaction, a $\nu_\mu$ gives rise to a $\mu^-$ and thus to a track,
 the $\nu_e$ yields an $e^-$ and thus an electromagnetic shower.

The early water Cherenkov detectors IMB \cite{imb}
and Kamiokande \cite{kamioka} reported 
anomalies in the ratio of muon to electron
neutrinos, while the tracking calorimeters NUSEX \cite{nusex}, 
 Frejus \cite{frejus}, and the Baksan \cite{baksan} scintillator detector
did not find any.

Later the Soudan 2 \cite{soudan}, MACRO \cite{macro} and 
SuperKamiokande \cite{skam} detectors
 reported deficits in the $\nu_\mu$ fluxes with
respect to the Monte Carlo (MC) predictions and a distortion of the angular
distributions; the $\nu_e$ distributions agree with MC. These features
may be explained in terms of $\nu_\mu \longleftrightarrow
\nu_\tau$ oscillations. A summary of the MACRO situation in 1998 is given
in ref. \cite{macro-1998}. \par Atmospheric neutrinos are well suited for 
the study
of neutrino oscillations, since they have energies from a fraction of GeV up 
to more than 100 GeV and they may travel distances $L$ from few tens of km 
up to 13000 km; thus $L/E_\nu$ ranges from $\sim 1$ km/GeV to $10^5$ 
km/GeV. Moreover one may consider that there are two identical sources for
a single detector: a near one (downgoing neutrinos) and a far one (upgoing
neutrinos). Atmospheric neutrinos are particularly suited to study oscillations
for small $\Delta m^2$. Finally, matter effects can be studied with high
energy atmospheric neutrinos. 

\section{Neutrino oscillations}
If neutrinos have non-zero masses, one has to consider 
the {\it weak flavour eigenstates}
$\nu_e,~\nu_\mu,~\nu_\tau$ and the {\it mass
eigenstates} $\nu_1,~\nu_2,~\nu_3$. 
The weak flavour eigenstates $\nu_l$ are linear combinations of the mass 
eigenstates $\nu_m$ through the elements of the mixing matrix $U_{lm}$:

\begin{equation}
\nu_l = \sum_{m=1}^3 U_{lm}\ \nu_m
\end{equation}
\noindent If the mixing angles are small, one would have $\nu_e \sim \nu_1$, 
 $\nu_\mu \sim \nu_2$, $\nu_\tau \sim \nu_3$;
 if they are large, the flavour eigenstates are well separated
 from those of mass.

In the simple case of only two flavour eigenstate neutrinos  
$(\nu_\mu,~\nu_\tau)$ which
oscillate with two mass eigenstates $(\nu_2,~\nu_3)$ one has

\begin{equation}
\left\{ \begin{array}{ll}
      \nu_\mu =~\nu_2 \cos\ \theta_{23} + \nu_3 \sin\ \theta_{23} \\
      \nu_\tau=-\nu_2\sin\ \theta_{23} + \nu_3\cos\ \theta_{23} 
\end{array} 
\right. 
\end{equation} 
\noindent where $\theta_{23}$ is the mixing angle.
 In this case one may easily compute the following expression for the 
survival probability of a $\nu_\mu$ beam:
{\footnotesize
\begin{equation}
P(\nu_\mu \rightarrow \nu_\mu) = 1- \sin^2 2\theta_{23}  
~\sin^2 \left( { {E_2-E_1}\over {2}} t \right) =
1- \sin^2 2\theta_{23}~\sin^2 \left( { 
{1.27 \Delta m^2 \cdot L}\over {E_\nu}} \right)
\end{equation}}
\noindent where $\Delta m^2=m^2_3-m^2_2$, $L$ is the distance travelled by the 
neutrino from production to detection. The probability 
for the initial $\nu_\mu$ to oscillate into a $\nu_\tau$ is:
\begin{equation}
P(\nu_\mu \rightarrow \nu_\tau) = 1 - P(\nu_\mu \rightarrow \nu_\mu) =
 \sin^2 2\theta_{23}~\sin^2 \left( { 
{1.27 \Delta m^2 \cdot L}\over {E_\nu}} \right)
\end{equation}
$\theta_{23}$ and $\Delta m^2$ may be determined 
from the variation of
$ P(\nu_\mu \rightarrow \nu_\mu) $ as a function of 
the zenith angle $\Theta$, or from the variation in $L/E_\nu$.

\section{Early experiments}

\begin{figure}
 \begin{center}
\mbox{\epsfig{figure=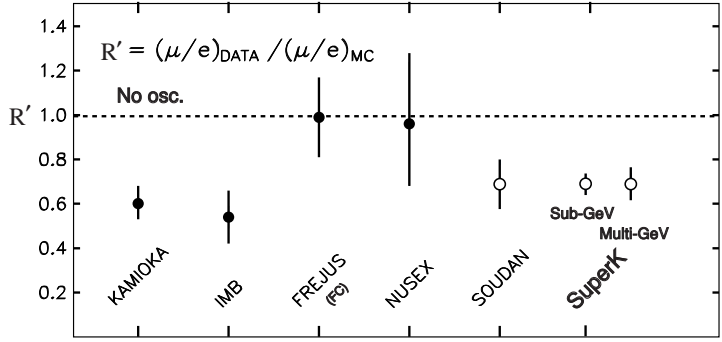,height=4.5cm}}
\caption {Double ratios $R^\prime$
measured by several atmospheric neutrino experiments.}
\label{fig:ratios}
\end{center}
\end{figure}

The early water Cherenkov detectors and the tracking calorimeters
measured $\nu_\mu$ 
and $\nu_e$ CC interactions. The results (see Fig. \ref{fig:ratios}) were
expressed in terms of the double ratio $R^\prime =R_{obs}/R_{MC}$, where
$R_{obs} = (N_{\nu_\mu} / N_{\nu_e})_{obs}$ is the ratio of observed 
$\mu$ and $e$ events and $R_{MC} = (N_{\nu_\mu}/ N_{\nu_e})_{MC}$ 
is the same ratio for Monte Carlo 
simulated events. While the single ratio  
$(N_{\nu_\mu})_{obs}/(N_{\nu_\mu})_{MC}$ is
affected by large theoretical and systematic uncertainties, in the 
double ratio $R^\prime$ most systematic uncertainties cancel.

The $R^\prime$ double ratios from IMB and Kamiokande
were smaller than expectations, while the
NUSEX and Frejus $R^\prime$ agreed with expectations.
 The Baksan scintillation telescope detected 
upthroughgoing muons arising from $\nu_\mu$ interactions in the
rock below the detector \cite{baksan}. The
average $\nu_\mu$ energy for these events is 
$50 \div 100$ GeV. They did not find deviations from the predictions 
in the total 
number of events, but found an anomalous angular distribution.

Later, the Soudan 2 tracking and shower calorimeter detector confirmed 
the anomaly
in the $\nu_\mu/ \nu_e$ double ratio for contained events \cite{soudan}, 
 Fig. \ref{fig:ratios}.

MACRO \cite{macro}  reported in 1995 a measurement
of upthroughgoing muons coming from $\nu_\mu$ of 
 $\langle E_\nu \rangle \sim 50$ GeV, 
 in which there was a global deficit in the total number of observed upgoing
muons and an 
anomalous zenith angle distribution. These features were later confirmed 
by the analyses of the whole samples of upthroughgoing muons \cite{macro} 
and of lower energy events ($\langle E_\nu \rangle \simeq 4$ GeV) \cite{macro}.

SuperKamiokande (SK) confirmed the anomalous double 
ratio, Fig. \ref{fig:ratios},
 and provided a wealth of informations for sub-GeV and multi-GeV 
$\nu_\mu,~\nu_e$ and for higher energy upthroughgoing muons and stopping 
muons \cite{skam}.

 After 1998 new results were presented
by the three experiments at most Conferences. Here we shall review 
their results. 

\section{Results from the Soudan 2 experiment}
The Soudan 2 experiment uses a modular fine grained tracking and showering
calorimeter of $963$ t. It is located $2100~ m.w.e.$ underground in the
Soudan Gold mine in Minnesota.
 Its overall dimensions are 8m$\times$16m$\times$5m. The detector 
is made of 224 modules of 1m$\times$1m$\times$2.5m. The
bulk of the mass consists of 1.6 mm thick corrugated  steel sheets 
 interleaved with drift tubes. The detector is 
surrounded by an anticoincidence  shield.

\begin{figure}
 \begin{center}
\mbox{\epsfig{figure=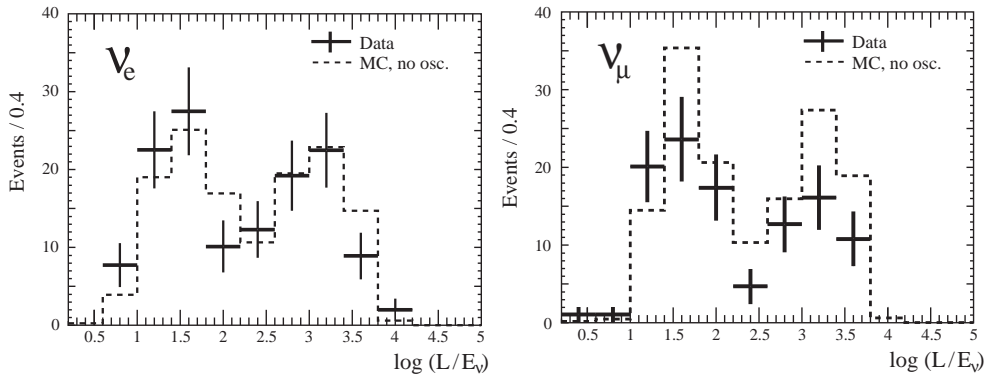,height=5cm}}
\caption {Distribution in $\log L/E_\nu$~ for $\nu_e$ and $\nu_\mu$ Soudan 2
CC HiRes events (black crosses) compared with the MC predictions for 
no oscillations (dashed lines). Only statistical errors are shown. The 
MC is rate-normalized to the $\nu_e$ data.}
\label{fig:sres}
\end{center}
\end{figure}

The neutrino contained events
are selected by a combination of a two-stage software
filter and a two-stage physicist scan. 
  Topologies for contained events include single track and
single shower events and multiprong events.  An
event having a leading, non-scattering track with ionization $dE/dx$
compatible with that from a muon is a candidate CC event of
$\nu_\mu$ flavour; an event yielding a relatively energetic shower 
is a candidate $\nu_e$ CC event. Multiprong events are not
considered at present. Events without hits in the 
shield 
are called {\it Gold Events}, while events with two or more
hits in the shield are called {\it Rock Events}.

\begin{table}
\begin{center}
\begin{tabular}
{cccc}\hline
 & Data Raw & Data Back. removed & MC \\ \hline
Track     & 133 & $105.1\pm 2.7$ & 193.1\\
Showers   & 193 & $142.3\pm 13.9$ & 179.0\\ \hline
\end{tabular}
\end {center}
\caption {Summary for the Soudan 2 Gold data.
A single track is due to a $\nu_\mu$, a single shower is due to a 
$\nu_e$. The
MC predictions were obtained using the $\nu$ Bartol flux applied to the 
Soudan 2 site and normalized to a 5.1 kt$\cdot$yr exposure.} 
\label{tab:soudan}
\end{table}

The sample of fully contained events consists mostly of quasi-elastic 
neutrino reactions, but include a
background of photons and neutrons from cosmic ray muon
interactions in the surrounding rock. 
 The track and shower events for a 5.1 kt$\cdot$yr exposure are summarized in
Table \ref{tab:soudan}, where they are compared with the MC 
predictions
based on the  Bartol neutrino flux \cite{Agrawal96}.
After corrections for cosmic ray muon induced background, the Soudan 2 
double ratio for the whole zenith angle range ($-1 \leq \cos \Theta \leq 1$) 
is $R^\prime=(N_\mu/ N_e)_{DATA}/ (N_\mu/ N_e)_{MC} 
= 0.68 \pm 0.11_{stat} \pm 0.06_{sys}$
which is consistent with muon neutrino oscillations, see 
Fig. \ref{fig:ratios}. \par

Selecting a high resolution (HiRes) sample
of events for which the resolution in $\log (L/E_\nu)$ is better than 
0.5, the flavour tagging is estimated to be correct for more than $92 \%$ of
the events.  After
background subtraction there are 106.3 events of $\nu_\mu$ flavour and 
132.8 events of $\nu_e$ flavour. Using these events, whose mean 
energy is slightly higher, the double ratio 
is $R = 0.67 \pm 0.12$. The $\log (L/E_\nu)$ distribution for $\nu_e$ and 
$\nu_\mu$ charged current events compared to the no oscillation MC
predictions normalized to the $\nu_e$ data is shown in Fig. \ref{fig:sres}. 
Notice that the $\nu_e$ data agree with the no oscillation MC 
predictions, while the $\nu_\mu$ data are lower; this is consistent with
oscillations in the $\nu_\mu$ channel and no oscillations for $\nu_e$. The 
double peak structure arises from the acceptance of the 
apparatus.  The 90\% C.L. allowed region in the 
$\sin^2 2\theta - \Delta m^2$ 
plane, computed using the Feldman-Cousins method \cite{feldman-cousins} 
is shown in Fig. \ref{fig:sk2}b, where it is compared with the allowed
regions obtained by the SK and MACRO experiments.

\section{Results from the MACRO experiment}
\begin{figure}
 \begin{center}
\mbox{\epsfig{figure=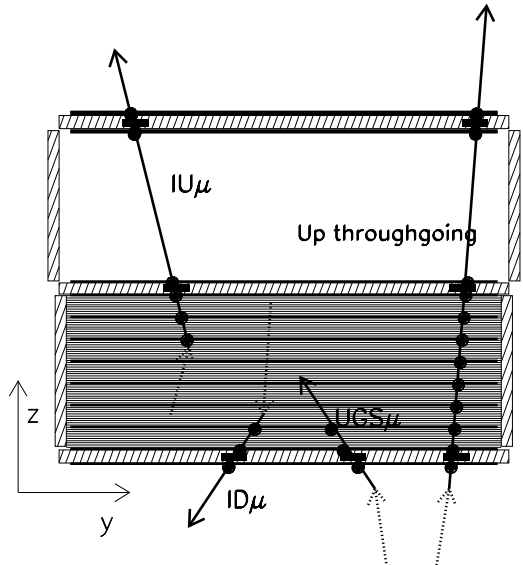,height=8cm}}
\vspace{-0.5cm}
\caption {Vertical section of the MACRO detector. Event 
topologies induced
by $\nu_\mu$ interactions in or around the detector. IU$_\mu=$ semicontained 
Internal Upgoing $\mu$; 
 ID$_\mu=$ Internal Downgoing $\mu$; UGS$_\mu=$ Upgoing Stopping 
$\mu$; Upthroughgoing =
upward throughgoing $\mu$. The black circles indicate the streamer tube 
hits and the black boxes the scintillator hits.}
\label{fig:topo}
 \end{center}
\end{figure}

The MACRO detector was located in Hall B of the Gran Sasso
Laboratory, at an average rock overburden of 3700 hg/cm$^2$.
It was a rectangular box,
76.6m$\times$12m$\times$9.3m, divided
longitudinally in six supermodules and vertically in a lower
part (4.8 m high) and an upper part (4.5 m high) \cite{macro}. The
detection elements were planes of streamer tubes for tracking
and liquid scintillation counters for 
the determination of the direction (versus) by the time-of-flight 
(T.o.F.) method. The lower half of
the detector was filled
with trays of crushed rock absorbers alternating with streamer tube
planes; the upper part was open and contained the
electronics. 
 Fig. \ref{fig:topo}, a vertical section of the detector, shows 
the different topologies of $\nu_\mu$ events.

In the MC simulation of upthroughgoing muons, the neutrino flux computed
by the Bartol group \cite{Agrawal96} and
the cross sections for the neutrino interactions calculated
using the deep inelastic  parton distribution \cite{Gluck95} are used.
 The propagation of muons to the detector was done using the
energy loss calculation in standard rock \cite{Lohmann85}. 
 For the low energy data, the simulations use the Bartol neutrino
flux and the low energy neutrino cross sections \cite{lipari94}.

\begin{table}
\begin{center}
\begin{tabular}{cccc}\hline
 & Events & MC$_{\mbox{no~osc}}$ & $R=$ Data/MC$_{\mbox{no~osc}}$ \\
Upthr.   & 809 & $1122\pm 190.7$ & $0.721 \pm 0.026_{stat} 
                   \pm 0.043_{sys} \pm 0.123_{th}$\\
IU      & 154 & $285\pm 28_{sys} \pm 71_{th}$ & $0.54 
                 \pm 0.04_{stat} \pm 0.05_{sys} \pm 0.13_{th}$ \\
ID+UGS  & 262 & $375 \pm 37_{sys} \pm  94_{th}$ & $0.70 
                 \pm 0.04_{stat} \pm 0.07_{sys} \pm 0.17_{th}$ \\
\hline
\end{tabular}
\end {center}
\caption {Summary of the MACRO events after background 
subtraction. For each topology the number of
measured events, the MC predictions for no oscillations and the ratio 
$R=$ Data/MC$_{\mbox{no~osc}}$ are given. Data and MC refer to the angular
region $-1 < \cos \Theta < 0$.} 
\label{tab:macro}
\end{table}

\begin{figure}
 \begin{center}
\mbox{\epsfig{figure=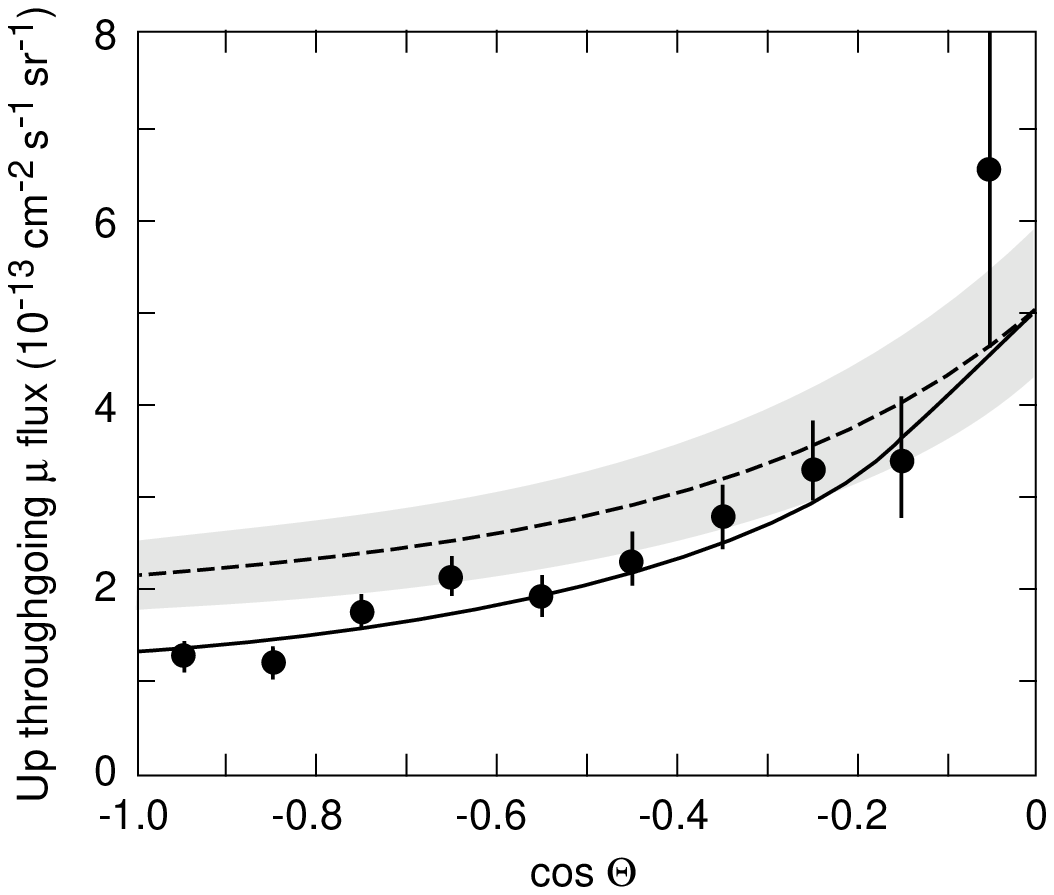,height=6.4cm}}
	\hspace{2mm}
\mbox{\epsfig{figure=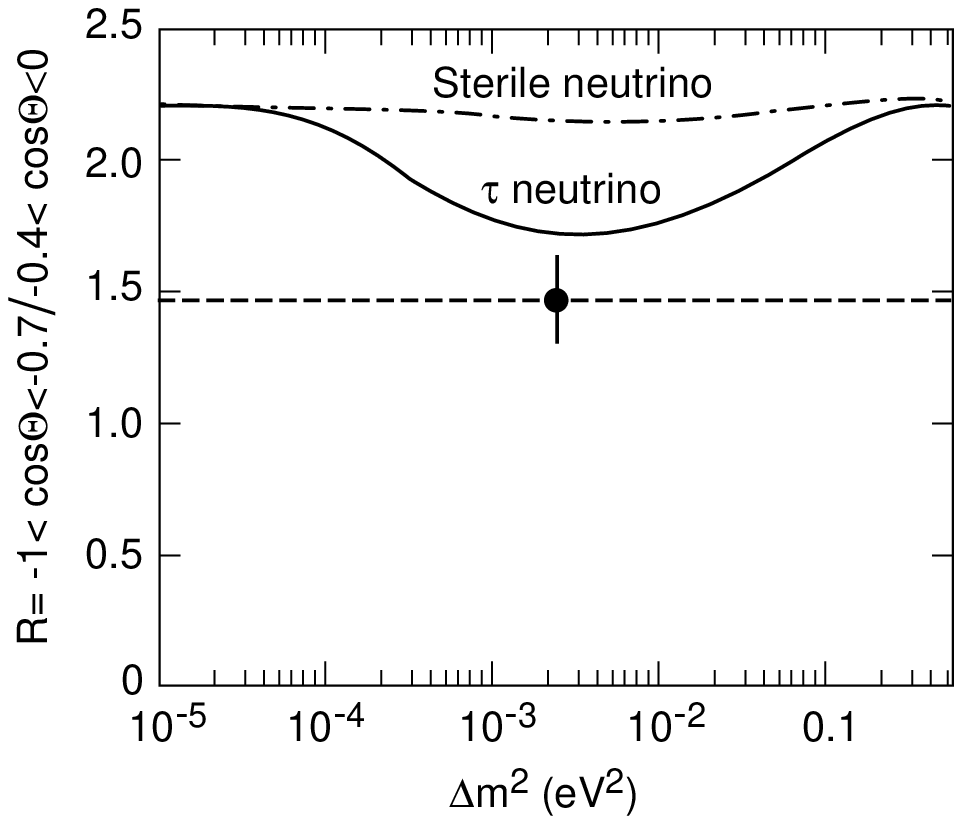,height=6.4cm,width=8.2cm}}
{\footnotesize \hspace{2.5cm} (a) \hspace {8cm} (b)}
\caption{(a) Zenith distribution of the upthroughgoing muons in 
MACRO. The 
data (black points) have error bars with statistical and systematic errors
added in quadrature. The shaded region shows the theoretical scale error 
band of $\pm 17\%$
on the normalization of the Bartol flux for no oscillations.
 The solid line is the fit to an oscillated flux which yields
 maximal mixing and $\Delta m^2 = 2.5 \cdot 10^{-3} $ eV$^2$. 
(b) Ratio of events with $-1< \cos \Theta < -0.7$ to events with 
$-0.4 <\cos \Theta < 0$ as a function of $\Delta m^2$ for maximal mixing.
 The black point with error bar is the measured value, the solid line is 
the prediction for
$\nu_\mu \longleftrightarrow \nu_\tau$ oscillations, the dashed-dotted line 
is the prediction for
$\nu_\mu \longleftrightarrow \nu_{sterile}$ oscillations.}
\label{fig:cos-uptr}
\end{center}
\end{figure}

\noindent {\bf \boldmath $\bullet$ Upthroughgoing muons ($E_\mu > 1$ GeV)}
 They come from  
interactions in the rock below the detector of $\nu_\mu$ with
$\langle E_\nu \rangle \sim 50$ GeV. The MC uncertainties arising from the 
neutrino flux, cross section and muon propagation on the
expected flux of muons are estimated to be $\sim$17\%; 
 this systematic uncertainty on the upthroughgoing muons flux is 
mainly a scale error. The  ratio of the observed
number of events to the expectation without oscillations 
in $-1 < \cos \Theta < 0$ is
$R=0.721 \pm 0.026_{stat} \pm 0.043_{sys} \pm 0.123_{th}$.
 Fig. \ref{fig:cos-uptr}a shows the zenith angle distribution of the measured
flux of upthroughgoing muons;
 the MC expectation for no oscillations is indicated by the dashed
line. The best fit to the number of events and to the shape of
the zenith angle distribution, assuming 
$\nu_\mu \longleftrightarrow \nu_\tau$ oscillations, yields
$\sin^2 2\theta =1$ and  $\Delta m^2= 2.5 \cdot 10^{-3}$ eV$^2$ (solid line).

The 90\% C.L. allowed region in the $\sin^2 2\theta - \Delta m^2$ 
plane, computed using the Feldman-Cousins method \cite{feldman-cousins} 
is shown in Fig. \ref{fig:sk2}b, where it is compared with those 
obtained by the SuperKamiokande and Soudan 2 experiments.

\noindent {\bf \boldmath $\bullet~\nu_\mu \longleftrightarrow \nu_\tau$ 
against $\nu_\mu \longleftrightarrow \nu_{sterile}$}
 
Matter effects due to 
the difference between the weak interaction effective potential for 
muon neutrinos with respect to sterile neutrinos, which have null
potential, yield different total number and different zenith 
distributions of upgoing muons. In Fig. \ref{fig:cos-uptr}b the measured ratio 
between the events with $-1 < \cos \Theta < -0.7$ and the events with 
$-0.4 < \cos \Theta < 0$ is shown \cite{macro}. In this ratio most 
of the theoretical uncertainties
on neutrino flux and cross sections cancel. The remaining  
error, $\leq 5$\%, comes from uncertainties on the kaon/pion 
fraction, on the cross sections
 of almost vertical and almost horizontal
events and on the seasonal variation of the 
ratio. The systematic experimental error on the ratio, due to analysis
cuts and detector efficiencies, is 4.6\%. Combining the experimental
and theoretical errors in quadrature, a global extimate of 6\% is obtained.
MACRO measured 305 events with $-1 < \cos \Theta < -0.7$ and 206 events with
$-0.4 < \cos \Theta <0$; the ratio is $R=1.48 \pm 0.13_{stat} \pm 0.10_{sys}$.
For $\Delta m^2=2.5 \cdot 10^{-3} $ eV$^2$ and maximal mixing, the minimum 
expected value of the ratio 
for $\nu_\mu \longleftrightarrow \nu_\tau$
is $R_\tau=1.72$ and for $\nu_\mu \longleftrightarrow \nu_{sterile}$ is
$R_{sterile}=2.16$. One concludes that 
$\nu_\mu \longleftrightarrow \nu_{sterile}$ oscillations (with any mixing) are 
excluded at 99\% C.L.
compared to the $\nu_\mu \longleftrightarrow \nu_\tau$ 
channel with maximal mixing and $\Delta m^2=2.5 \cdot 10^{-3} $ eV$^2$.

\noindent {\bf \boldmath $\bullet$ $\nu_\mu$ energy estimate by 
Multiple Coulomb Scattering of muons} 

Since MACRO was not equipped with a magnet, 
 the only way to estimate the muon energy is through Multiple Coulomb 
Scattering (MCS) of muons in the absorbers. The
r.m.s. of the lateral displacement of a relativistic muon  travelling 
for a distance $X$ can be written as $\sigma_{MCS} \sim \frac {X} {p_\mu} \cdot
\sqrt{\frac{X}{X^0}}$ , where $p_\mu$ (GeV/c) is the muon momentum and 
$X/X^0$ is the amount of crossed material in terms of radiation 
lengths \cite{miriam-mcs,spurio-mcs}. A muon crossing the whole
apparatus, on the vertical, has $\sigma_{MCS} \simeq 10$ cm/$E_\mu$(GeV). The 
muon energy estimate can be performed up to a saturation point, occurring when
$\sigma_{MCS}$ is comparable with the detector space resolution. 

Two analyses were performed. The first  
was made studying the deflection of upthroughgoing muons 
using the streamer tubes in digital mode \cite{spurio-mcs}. This method 
has a spatial
 resolution of $\sim 1$ cm. The data were divided into 3
subsamples with different average energies, in 2 subsamples in zenith angle
and finally in 5 subsamples with different average values of $L/E_\nu$. 

\begin{figure}
 \begin{center}
\mbox{\epsfig{figure=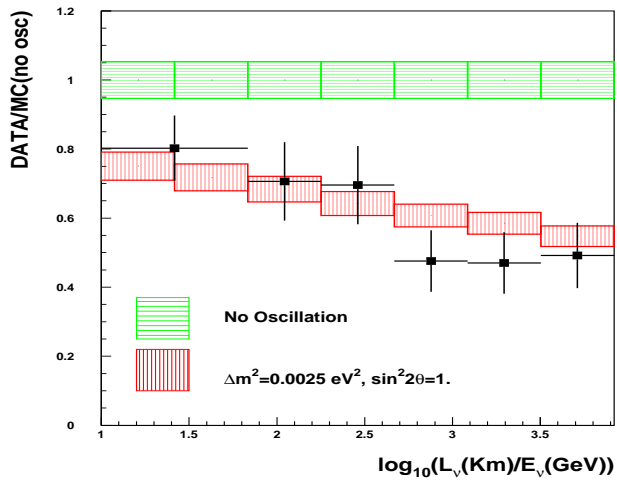,height=6.7cm,width=8.5cm}}
\vspace{-3mm}
\caption {Data/MC$_{\mbox{no~osc}}$ as a function of $\log_{10} (L/E_\nu)$
obtained from upthroughgoing muon analysis in MACRO $^{10}$. The black 
points are the data, the shaded regions are
the MC predictions for $\nu_\mu \longleftrightarrow \nu_\tau$ oscillations
and for the no oscillation hypothesis.} 
\label{fig:mcs}
\end{center}
\end{figure}

The second analysis was performed
using the streamer tubes in ``drift mode" \cite{miriam-mcs}; the time 
information was given by a TDC. The space resolution is $\simeq 3$ mm. 
 For each muon, 7 MCS sensitive variables were defined and given in input 
to a Neural 
Network (NN), previously trained with MC events of known  energy 
crossing the detector at different zenith angles. The NN output allows to
separate the sample of upthroughgoing muons in 4 subsamples with 
average energies $E_\mu$ of 12, 20, 50 and 100 GeV. The ratios 
Data/MC$_{\mbox{no~osc}}$ as a function of 
$\log_{10} (L/E_\nu)$ obtained from upthroughgoing muons are plotted in 
Fig. \ref{fig:mcs}; they are in agreement with the 
$\nu_\mu \longleftrightarrow \nu_\tau$ oscillation hypothesis. 

\noindent {\bf \boldmath $\bullet$ Low energy events} 

\begin{figure}
 \begin{center}
\mbox{\epsfig{figure=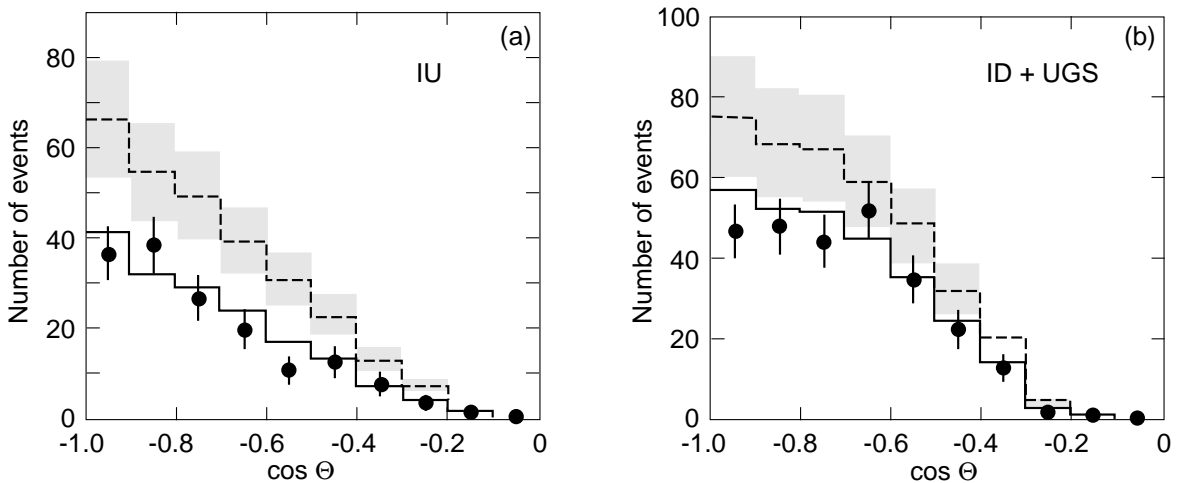,height=6.5cm}}
\caption {(a) Measured zenith distributions for the IU events 
 and (b) for the
ID+UGD events in MACRO (black points). The shaded regions 
correspond to MC predictions assuming no oscillation. The full line is 
the expectation for $\nu_\mu \longleftrightarrow \nu_\tau$ oscillations with 
maximal mixing and $\Delta m^2 =2.5\cdot 10^{-3}$ eV$^2$.}
\label{fig:low_cosze}
\end{center}
\end{figure}

{\it Semicontained upgoing muons} (IU) come from
$\nu_\mu$ interactions inside the lower apparatus. Since two
scintillation counters are intercepted, the T.o.F. method
is applied to identify  upward going muons. The average 
parent neutrino energy for these events is $\sim 4$ GeV. 

{\it Up stopping muons} (UGS)
are due to external $\nu_\mu$ interactions yielding upgoing muons 
stopping in the detector; the {\it semicontained downgoing muons} (ID) are 
due to 
downgoing $\nu_\mu$'s with interaction vertices in the lower MACRO.
The events are found by  topological criteria; the lack
of time information prevents to distinguish between the two subsamples.
An almost equal number of UGS and ID 
events is expected.

 The number of events and the angular distributions are compared with the MC
predictions without oscillations in Table \ref{tab:macro} and 
Fig. \ref{fig:low_cosze}. The low
energy data show a uniform deficit of the measured number of events
over the whole angular distribution with respect to the predictions;
the data are in good agreement with the predictions based on 
$\nu_\mu \longleftrightarrow \nu_\tau$ 
oscillations with the parameters obtained from the upthroughgoing muon
sample.

\section{Results from the SuperKamiokande experiment}

SuperKamiokande \cite{skam} (SK) is a large cylindrical water 
Cherenkov detector 
of 39 m diameter and 41 m height containing 50 kt of 
water (the fiducial mass of the detector for atmospheric neutrino analyses 
is 22.5 kt); it is seen by 11146, 50-cm-diameter 
inner-facing phototubes. The 2 m thick outer layer of water acts 
as an anticoincidence and is seen by 1885 smaller outward-facing
photomultipliers. The ultra pure water
has a light attenuation of almost 100 m. The detector 
is located in the Kamioka mine, Japan, under $2700~ m.w.e.$

Atmospheric neutrinos are detected in SK
by measuring the Cherenkov light
generated by the charged particles produced in the neutrino
CC interactions with the water nuclei.
 Thanks to the high PMT coverage, the experiment is characterised by a 
good light yield ($\sim 8$ photo-electrons per
MeV) and can detect events of energies as low as $\sim
5$ MeV.

\begin{figure}
 \begin{center}
\mbox{\epsfig{figure=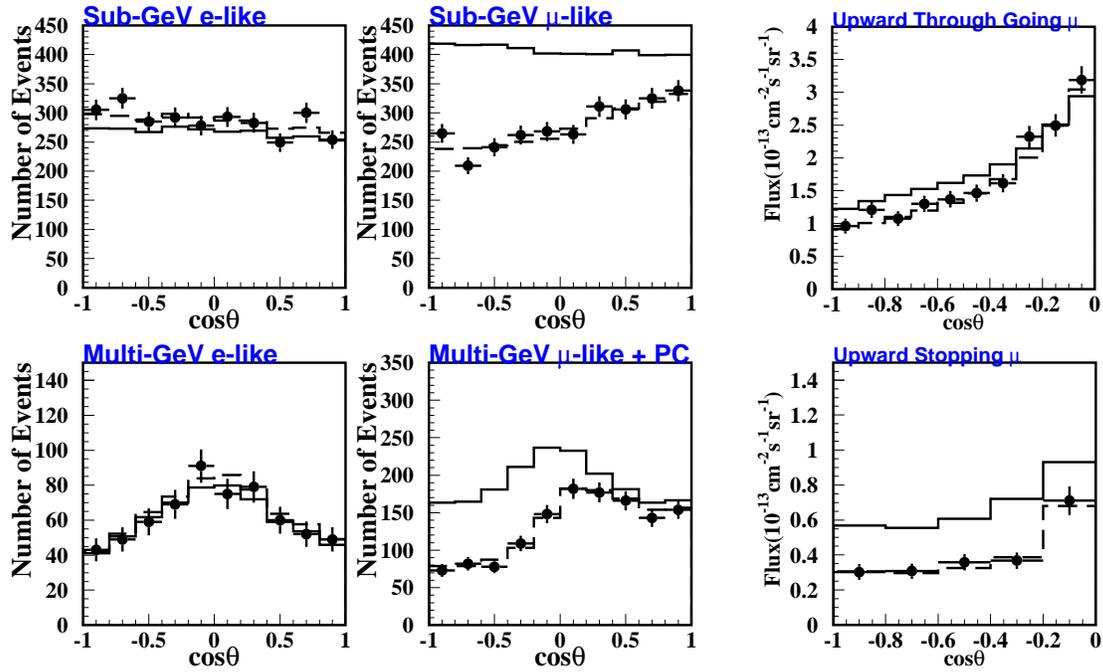,height=9cm}}
\caption{Zenith distributions for
  SK data (black points) for $e$-like and $\mu$-like sub-GeV and multi-GeV 
events and for throughgoing and stopping muons. The
solid lines are the no oscillation MC predictions, the dashed
lines refer to $\nu_\mu\longleftrightarrow\nu_\tau$
oscillations with maximal mixing and $\Delta m^2=2.5\cdot 10^{-3}$ eV$^2$.}
\label{fig:sk3}
\end{center}
\end{figure}

 The large detector mass and the possibility of clearly
defining a large inner volume allow to collect a high
statistics sample of {\it fully contained} events (FC) up to relatively high
energies (up to $\sim 5$ GeV). The FC events have both the
neutrino vertex and the resulting particle tracks entirely within the
fiducial volume; they yield rings of Cherenkov light on the 
PMTs. The contamination from downward-going
cosmic muons is drastically
reduced by the containment requirement on the primary vertex coordinates.
 Fully contained events can be further subdivided
into two subsets, the so-called {\it sub-GeV} and {\it multi-GeV} events,
with energies below and above 1.33 GeV, respectively.
In SK jargon FC events include only single-ring events,
while {\it multi-ring} ones (MRING) are treated as a separate category.
Another sub-sample, defined as the {\it partially contained} events (PC), is
represented by those CC interactions where the vertex
is still within the fiducial volume, but at least a primary charged 
particle, typically
the muon, exits the detector without releasing all of its energy. In this case
the light pattern is a filled circle.
 For these events the energy resolution is worse than
for FC interactions. {\it Upward-going muons} (UPMU), produced by 
neutrinos coming from
below and interacting in the rock, are further subdivided into 
{\it stopping  muons} ($\langle E_\nu \rangle \sim 7$ GeV) and 
{\it throughgoing muons} ($\langle E_\nu \rangle \sim 70 \div 80$ GeV), 
 according to whether or not they stop in the
detector. The samples defined above explore different
ranges of 
neutrino energies \cite{engel_gaisser_stanev}.

\begin{table}
\begin{center}
\begin{tabular}
{cccc}\hline
 & Data & MC no osc. & $\langle E_\nu \rangle$ (GeV) \\ \hline
Sub-GeV $e$-like   & 2864 & 2668 & 0.6\\ \hline
Sub-GeV $\mu$-like & 2788 & 4073 & 0.6\\ \hline
Multi-GeV $e$-like & 626 & 613 & 3\\ \hline
Multi-GeV $\mu$-like & 558 & 838 & 3\\ \hline
\end{tabular}
\end {center}
\caption {Summary of the SK data. For $e$-like and $\mu$-like
Sub-GeV and Multi-GeV events, the table gives the number of measured 
events, the MC predictions for no oscillations and the approximate 
average energy.} 
\label{tab:superk}
\end{table}

Particle identification in SuperKamiokande is performed using
likelihood functions to parametrize the sharpness of
the Cherenkov rings, which are more diffused for electrons than for
muons. The algorithms are able to discriminate
the two flavours with high purity
(of the order of $98\%$ for single track events). 
The zenith angle distributions for $e$-like and $\mu$-like 
sub-GeV and multi-GeV events are shown in Fig. \ref{fig:sk3}.
 The electron-like events are in 
agreement with the MC predictions in absence of oscillations,
 while the muon data are lower than the no oscillation
expectations. The number of measured and expected events are summarized 
in Table \ref{tab:superk}. Moreover, the $\mu$-like data exhibit an up/down 
asymmetry in zenith angle, while no significant asymmetry is observed 
in the $e$-like data \cite{skam}. 

\begin{figure}
 \begin{center}
\mbox{\epsfig{figure=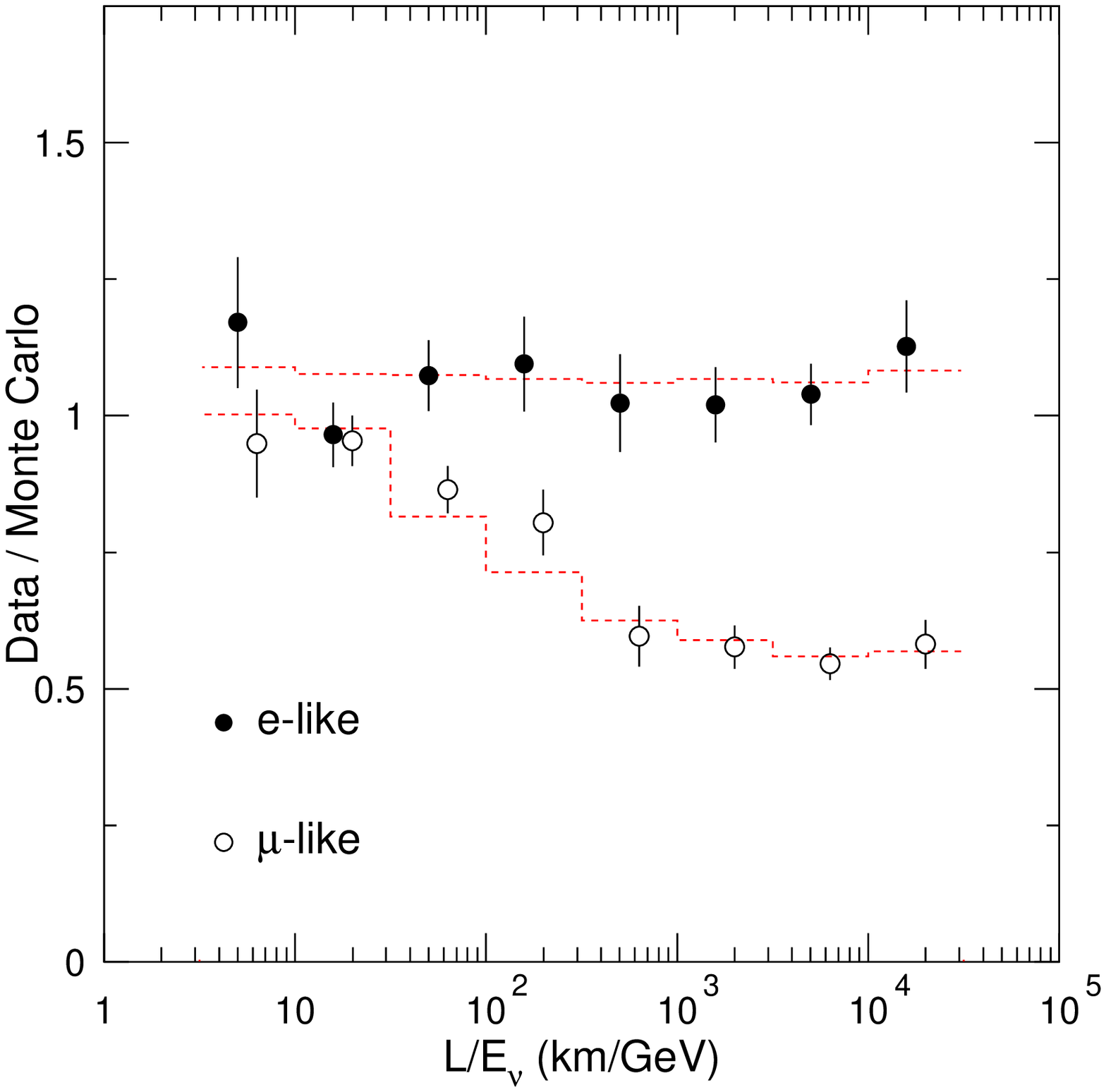,height=6.5cm}}
	\hspace{2mm}
\mbox{\epsfig{figure=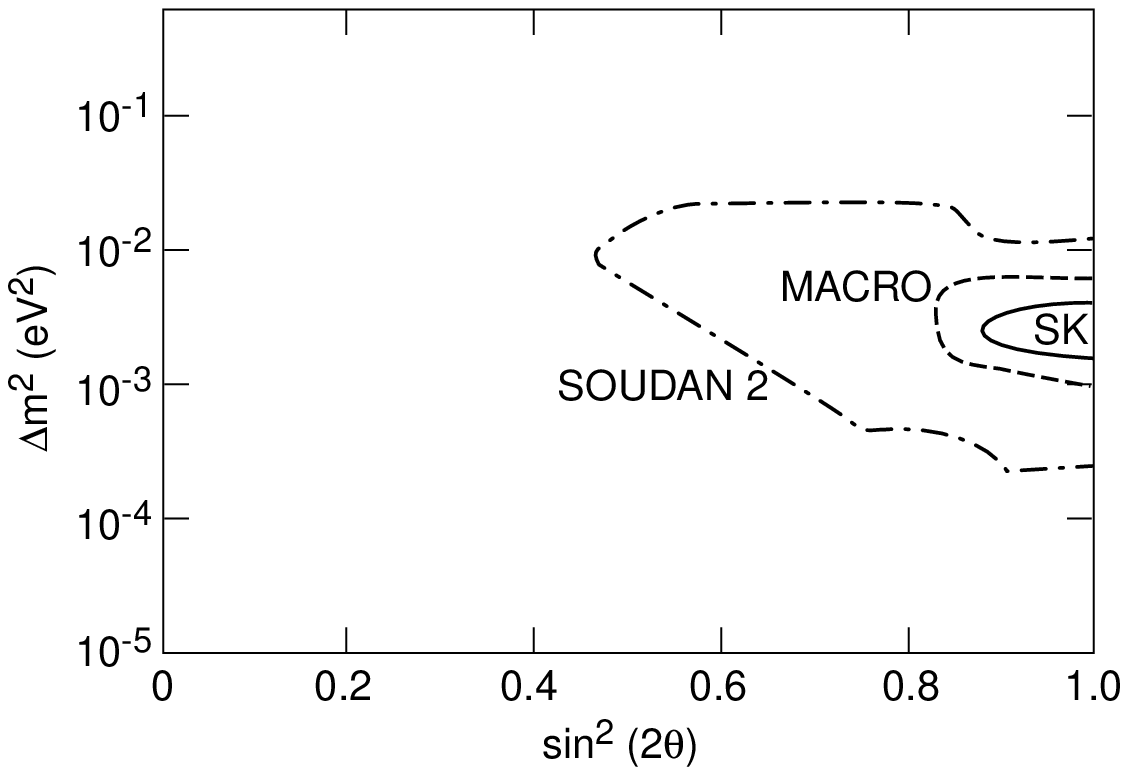,height=6.7cm,width=8.5cm}}
{\footnotesize \hspace{2.5cm} (a) \hspace {8cm} (b)}
\caption{(a) SK ratios between observed and expected numbers of $e$-like and
$\mu$-like events as a function of $L/E_\nu$. 
(b) 90\% C.L. allowed region contours for $\nu_\mu \longleftrightarrow \nu_\tau$
oscillations obtained by the SuperKamiokande, MACRO and Soudan 2 experiments.}
\label{fig:sk2}
\end{center}
\end{figure}

The recent 
value for the double ratio $R^\prime$ reported by 
SK, based on 1289 days of data, is
$0.638^{+0.017}_{-0.017}\pm 0.050$ for the sub-GeV sample and
$0.675^{+0.034}_{-0.032}\pm 0.080$ for the multi-GeV sample
(both FC and PC), Fig. \ref{fig:ratios}.
The ratio between observed and expected numbers of $e$-like and
$\mu$-like events as a function of $L/E_\nu$ is shown in Fig. \ref{fig:sk2}a.
 The ratio $e$-like events/MC do not depend from $L/E_\nu$ 
 while $\mu$-like events/MC show a
dependence on $L/E_\nu$ consistent
with the oscillation hypothesis. Interpreting the muon-like event deficit 
as the result of $\nu_\mu \longleftrightarrow \nu_\tau$
oscillations in the two-flavour mixing scheme, SuperKamiokande computes
an allowed domain for the oscillation parameters \cite{skam}, see 
Fig. \ref{fig:sk2}b.
 The events are binned in a multi-dimensional space defined by particle
type, energy and zenith angle, plus a set of parameters to account for
systematic uncertainties. The best fit using FC, PC, UPMU and MRING events
\cite{skam} corresponds to maximal mixing and $\Delta m^2 =
2.5 \cdot 10^{-3}$ eV$^2$, Fig. \ref{fig:sk2}b.

SK reported also data on upthroughgoing muons, which agree 
with the predictions of an oscillated flux with the above 
parameters. Notice that the average energies are larger than those of
the corresponding sample in MACRO.

\begin{figure}
 \begin{center}
\mbox{\epsfig{figure=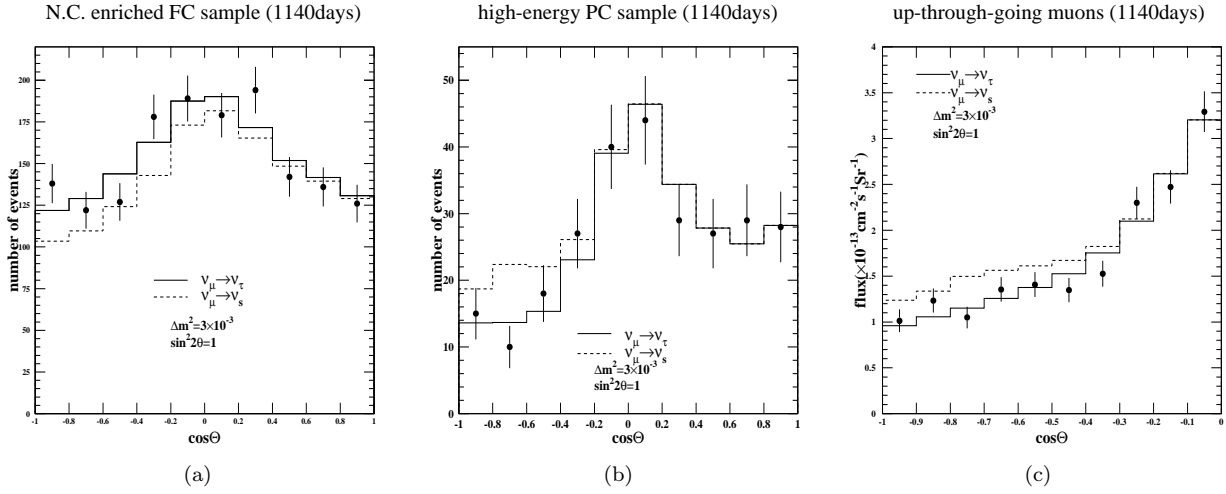,height=6cm}}
{\footnotesize  (a) \hspace{5cm} (b) \hspace{5cm} (c)}
\caption{SK data. Zenith angle distributions of (a) NC enriched sample,
 (b) high-energy PC muon sample, (c) upthroughgoing muon sample.}
\label{fig:zen_tau_s}
\end{center}
\end{figure}

{\bf \boldmath $\bullet~\nu_\mu \longleftrightarrow \nu_\tau$ against
$\nu_\mu \longleftrightarrow \nu_{sterile}$} 
 
If the observed deficit of $\nu_\mu$ were due to $\nu_\mu\longleftrightarrow
\nu_{sterile}$ oscillations, then the number of events produced via
neutral current (NC) interaction for up-going neutrinos should also be reduced.
Moreover, in the case of $\nu_\mu\longleftrightarrow\nu_{sterile}$ 
oscillations, matter effects will
suppress oscillations in the high energy ($E_\nu > 15$ GeV) region.
The following data samples were used to search for these effects:
(a) NC enriched sample, (b) the high-energy ($E> 5$ GeV) PC sample and
(c) upthroughgoing muons.
The zenith angle distribution for each sample is shown in
Fig. \ref{fig:zen_tau_s}. The hypothesis test is performed using the
up($-1< \cos\Theta<-0.4$)/down($0.4 < \cos\Theta < 1$) ratio in 
samples (a) and (b) and
the vertical($-1 < \cos\Theta < -0.4$)/horizontal($-0.4 <\cos\Theta <0$) 
ratio in 
sample (c). The excluded regions obtained by a combined
((a),(b)and(c)) analysis and by the analysis of 1-ring-FC 
 show that $\nu_\mu\longleftrightarrow\nu_{sterile}$ oscillations are disfavored
with respect to $\nu_\mu\longleftrightarrow\nu_{\tau}$ oscillations at a C.L.
of 99\% \cite{skam}.

\section{Conclusions and outlook}
The results on atmospheric neutrinos obtained by the Soudan 2, MACRO
and SuperKamiokande experiments was discussed.
 The zenith angle distributions and the observed number of neutrino-induced 
muons disagree with the
predicted values for the no oscillation hypothesis.
For $\nu_e$-induced electrons there is no strong deviation from 
prediction. The ratio of muons to electrons normalized to the respective
MC predictions enhances the anomaly.
All muon data are in agreement with the hypothesis of two flavour
$\nu_\mu \longleftrightarrow \nu_\tau$ oscillations, with maximal mixing and
$\Delta  m^2 \sim 2.5 \cdot 10^{-3} $ eV$^2$. The  hypothesis of 
$\nu_\mu \longleftrightarrow \nu_{sterile}$ oscillations  is disfavoured at 
99\% C.L. for any mixing.
The 90\% C.L. contours of Soudan 2, MACRO and SuperKamiokande 
overlap, see Fig. \ref{fig:sk2}b.

The above  experiments on atmospheric neutrinos are disappearance 
experiments; future atmospheric $\nu$ experiments are under 
study (see for example ref. \cite{future}); also long baseline experiments 
using $\nu_\mu$ from accelerators are planned (see the lectures of ref. 
\cite{dilella}). The
 main goals of these experiments are the detection of the first
oscillation in $L/E_\nu$ and an appearance of $\nu_\tau$, to really prove the
oscillation hypothesis. They should also yield improved measurements of the
oscillation parameters.
Other goals include further discrimination of
$\nu_\mu \longleftrightarrow \nu_\tau$ from 
$\nu_\mu \longleftrightarrow \nu_{sterile}$
oscillations, and the detection of a possible small 
$\nu_\mu \longleftrightarrow \nu_e$ contribution. Eventually one would 
like a complete information on the $3\times 3$ oscillation matrix. 
 Matter effects could be
used to constrain hybrid oscillation models \cite{future} and measure the 
sign of $\Delta m^2$.

{\bf \boldmath $\bullet$ Water Cherenkov detectors.} 
The principle of a 650 kt (450 kt fiducial) water Cherenkov detector,
tentatively called UNO \cite{future}
(Ultra underground Nucleon decay and neutrino Observatory),
is currently being discussed. It could be 20-times SK; it would 
therefore increase 20 times the statistics and, in addition, it could
allow the study of $\nu_\mu \longleftrightarrow \nu_\tau$ oscillations and give
 an explicit $\tau$ appearance signal from atmospheric neutrinos. 

The AQUA-RICH \cite{future} project would be based on the Ring Imaging 
Cherenkov (RICH) technique. 
 This detector could reconstruct the neutrino interaction vertex and the 
particle flight direction, while the space-time structure of the
Ring Image would yield a measurement of the particle momentum
through multiple Coulomb scattering. A momentum resolution of 7\% could 
be achieved.
 This would yield an $L/E_\nu$ resolution sufficient to resolve the 
oscillation pattern.

{\bf \boldmath $\bullet$ Magnetized tracking calorimeters.} 
The advantage of such detectors is that they can measure the muon
charge, and thus separate the neutrino and antineutrino components.
 This should be useful to study matter effects, which can be 
significantly different in the two cases.
Furthermore, the energy of the hadronic system and the momentum of
semicontained muons could be measured accurately. 
MONOLITH \cite{future} (Massive Observatory for 
Neutrino Oscillations or LImits on
THeir existence) is a plan for
a 34 kt magnetised iron detector dedicated to the measurement of
atmospheric neutrinos. Its fiducial mass of about 26 kt would match the
mass of SK, while its superior $L/E_\nu$ resolution could allow the
reconstruction of the oscillation pattern.

\section{Acknowledgements}
We would to thank many colleagues for providing informations and
suggestions; in particular we acknowledge the cooperation  
of D. Bakari, G. Battistoni, Y. Becherini, 
 P. Bernardini, T. Montaruli, F. Ronga, E. Scapparone, M. Sioli, 
 A. Surdo.

\end{document}